\documentclass[a4paper,aps,prb,twocolumn]{revtex4}
\usepackage{amsmath}
\usepackage{graphicx}
\usepackage{subfigure}

\DeclareMathOperator{\arcsinh}{arcsinh}

\DeclareMathOperator{\arctanh}{arctanh}
\def\be{\begin{equation}}
\def\ee{\end{equation}}

\begin{document}

\title{Effect of inhomogeneous coupling on superconductivity}
\author{Yue Zou${}^1$, Israel Klich${}^2$, and Gil Refael${}^1$}
\affiliation{${}^1$Department of Physics, California Institute of
Technology, Pasadena, CA 91125\\ ${}^2$Kavli Institute for
theoretical physics, University of California Santa Barbara, CA
93106}
\date{\today}

\begin{abstract}
We investigate the influence of inhomogeneity in the pairing coupling
constant $U(\vec r)$ on dirty BCS superconductors, focusing on $T_c$, the order parameter
$\Delta(\vec r)$, and the energy gap $E_g(\vec r)$.
Within mean-field theory, we find that when the length-scale of the
inhomogeneity is comparable to, or larger than the coherence
length, the ratio $2E_g/T_c$ is significantly reduced from that of
a homogeneous superconductor, while in the opposite limit this
ratio stays unmodified. In two dimensions, when strong phase
fluctuations are included, the Kosterlitz-Thouless temperature
$T_{KT}$ is also studied. We find that when the inhomogeneity
length scale is much larger than the coherence length,
$2E_g/T_{KT}$ can be larger than the usual BCS value. We use our results to qualitatively
explain recent experimental observation of a surprisingly
low value of $2E_g/T_c$ in thin films.
\end{abstract}
\maketitle

\section{Introduction}

The presence of disorder in essentially all superconducting systems
makes research of the interplay of disorder and superconductivity
essential. In their pioneering work, Anderson\cite{Anderson1959}, and Abrikosov and Gorkov\cite{AG},
claimed that nonmagnetic impurities have no considerable effect on the
thermodynamic properties of s-wave superconductors; this result is
known as "Anderson theorem" for weakly disordered dirty
superconductors. Since the discovery and elucidation of the localization
phenomenon \cite{localization}, corrections to the Anderson theorem have been
intensively investigated
both experimentally\cite{Beasley1984,Dynes1986a,Dynes1986b,Dynes1989,Jaeger}
and
theoretically\cite{Maekawa1981,
Anderson1983,MaLee1985,Kapitulnik1985,Ramakrishnan,Finkelstein1987,Finkelstein1994,Larkin1999,Trivedi2001,Dubi2007}.
Within mean field theory, it has been shown that if one neglects
Coulomb interactions, pairing survives below the mobility edge until the
localization length reaches a critical
value\cite{MaLee1985,Kapitulnik1985}. But interactions
change this picture significantly, since the effect
of Coulomb repulsion is strengthened by localization, resulting in a
suppressed effective attractive
interaction and thus a reduced mean-field
 $T_c$\cite{Maekawa1981,Anderson1983,Ramakrishnan,Finkelstein1987,Finkelstein1994}.
An underlying assumption of these works is the uniformity of the
superconducting order parameter, which has been questioned by
numerical simulations in recent years \cite{Trivedi2001,Dubi2007}.

Experiments in this field focused on two-dimensional (2d)
superconductors, namely superconducting thin films. The disorder
in superconducting films is expected to reduce the superfluid
density and the phase ordering temperature, i.e., the
Kosterlitz-Thouless temperature $T_{KT}$, in addition to
suppressing the mean field $T_c$. These considerations naturally
lead to the possibility of a quantum superconductor-insulator
transition (SIT) at a critical amount of disorder or magnetic
field. Furthermore, the scale invariant nature of a film's
resistance raised expectations that such an SIT would exhibit many
universal features \cite{FisherGrinsteinGirvin}. The
superconducting-insulator transition was intensively studied
experimentally\cite{Haviland1989,Hebard1990,Hebard1992,Valles1992,Liu1993,Hsu1995,Valles1994,
Yazdani1995,Hsu1998,Goldman1998,Goldman2003,shahar2004}. The
theoretical viewpoint on these transitions took two main forms:
the nature of the SIT was interpreted either as the breaking of
Cooper pairs caused by amplitude
fluctuation\cite{Finkelstein1987,Finkelstein1994,Larkin1999}, or
localization of Cooper pairs resulting from phase
fluctuation\cite{FisherLee,Fisher1990,WenZee,FisherGrinsteinGirvin,Cha1991,Wallin1994}.
While the nature of the SIT in various systems is still debated,
in recent years the interest in this problem is further
intensified by the observation of a possible metallic phase
intervening the superconducting and insulating
phase\cite{Kapitulnik1996,Kapitulnik1999,Kapitulnik2001,Yoon2006,Merchant}.
This observation stimulated several theoretical proposals
\cite{vortexmetal,Phillips2001,Spivak2001,dissipation,auerbach,meir},
but its origin is still a mystery.

Motivated by the thin-film physics, more experimental studies
focused on the nature of the density of states (DOS) and the
quasi-particle energy gap of disordered single layer
superconducting thin
films\cite{Dynes1986a,Dynes1986b,Dynes1989,Valles1992,Valles1994,Hsu1995,Hsu1998}
and superconductor - normal-metal (SN) bilayers
\cite{Merchant,subgap,superweak}. Interestingly, these studies
found a broadening of the BCS peak and also a subgap density of
states\cite{
Dynes1986b,Valles1992,Valles1994,Hsu1995,Hsu1998,subgap}. Of
particular interest to us is the work in Ref. \onlinecite{superweak}, which studied a thin SN bilayer system,
and found a surprisingly low value of the ratio of the energy gap
to $T_c$, in contradiction to standard BCS theory, and  the theory
of proximity \cite{Cooper,deGennes,FF} where it is claimed that
the energy gap-$T_c$ ratio should be bounded from below by
$\sim3.52$. A drop below this bound, $2E_g/T_c<3.52$, was also
observed in amorphous Bi films as it approaches the disorder tuned
SIT \cite{Valles1992,Valles1994}. Similar trends were also
observed in SN bilayers in Ref. \onlinecite{Merchant} and in
amorphous tin films in Ref. \onlinecite{Dynes1989}.

In this paper we show that a reduction of the $2E_g/T_c$
ratio in a dirty superconductor could be explained as a consequence of inhomogeneity in the
pairing interaction. In SN bilayer thin films, thickness fluctuations
of either layer result in effective pairing inhomogeneity (in thin SN bilayers the effective
pairing is the volume averaged one, c.f., Ref. \onlinecite{deGennes,FF}
and Sec. \ref{sec3}). Such inhomogeneities in other systems occur due to grain boundaries, dislocations, or
compositional heterogeneity in alloys\cite{metals}.
For simplicity we will assume in our analysis that the pairing coupling constant
takes a one-dimensional modulating form:
\be
U(\vec r)=\bar U+U_Q\cos(Qx).\label{eq0}
\ee

In bilayer SN films, the effect of localization and
Coulomb interaction is minor compared to proximity effect,
and therefore we will neglect these complications in this work.

In our results, the ratio between the inhomogeneity length, $L\equiv
1/Q$, and the superconducting coherence length $\xi$, plays a
crucial role. When $Q\xi\gg 1$, the superconducting properties are
determined by an effective coupling $\bar U\lesssim U_{eff}<\bar
U+U_Q$ \cite{Podolsky}. In this limit, the ratio $2E_g/T_c$ is
preserved at the standard BCS value $\sim3.52$. Small corrections
are obtained when $1/(Q\xi)$ is finite. In the opposite limit,
$Q\xi\ll 1$, the system tends to be determined by the local value
of $U(x)$. Within mean field theory, the ratio $2E_g/T_c$ is
generally suppressed from the BCS value $3.52$; in 2d, however,
when one includes the thermal phase fluctuation and studies the
Kosterlitz-Thouless temperature, $T_{KT}$, the ratio $2E_g/T_{KT}$
can be larger than the usual BCS value. These results on
$2E_g/T_c$ are summarized in FIG. \ref{result}.

Our analysis is inspired by similar previously studied models.
Particularly, the $T_c$ of the clean case of this model has been
analyzed in Ref. \onlinecite{Podolsky}. Here we extend the study of
non-uniform pairing to both $T_c$ and zero-temperature
properties of disordered films, in the regime where the electron mean
free path $l$ obeys  $1/k_F\ll l\ll\xi_0\sim\frac{\hbar
  v_F}{T_c}$, which is relevant to the experiments of Long et al.\cite{subgap,superweak}.
  Note that while Anderson theorem
states that the critical temperature and gap of a homogenous
superconductor do not depend on disorder\cite{Anderson1959}, in an inhomogeneous
system the theorem does not hold. Indeed, we find that the results
of Ref. \onlinecite{Podolsky}, are modified in the dirty case. In
another related work, a system with a Gaussian distribution of the
inverse pairing interaction was studied
\cite{Larkin1972,Simons2001}. It was shown that an exponentially
decaying subgap density of states appears due to mesoscopic
fluctuations which lie beyond the mean field picture. Finally,
inhomogeneous coupling in the attractive Hubbard model
\cite{Dagotto2006} and lattice XY model \cite{Loh} were also
analyzed, with relevance to High-$T_c$ materials.

This paper is organized as follows. In Sec. \ref{sec2} we review
the quasiclassical Green's function formalism which we use, and briefly demonstrate
how it works for the usual dirty superconductors with spatially
uniform coupling constant. Then, in Sec. \ref{inhomo} we discuss the cases with nonuniform
coupling classified by the competition of two length scales: the
coherence length $\xi$ and the length scale associated with the
variation of the coupling constant $L=1/Q$. We will also discuss the
effect of other types of inhomogeneities briefly. In section \ref{sec3} we
provide a useful analogy with superconductor-normal metal
superlattice to provide more physical intuition about our results on
the energy gaps. In section \ref{summary} we will summarize our analysis and
discuss the connection with experimental results.

\section{The gap equation of a nonuniform film \label{sec2}}

The starting point of our analysis is the standard s-wave BCS Hamiltonian:
\begin{eqnarray}\label{1}
H&=&H_0+H_{int}+H_{imp},\nonumber\\
H_0&=&\sum_{\sigma}\psi^{\dag}_{\sigma}(\vec
r)\hat{\xi}\psi(\vec r)_{\sigma},\nonumber\\
H_{int}&=&-U(\vec r)\psi^{\dag}_{\downarrow}(\vec
r)\psi^{\dag}_{\uparrow}(\vec r)\psi_{\uparrow}(\vec
r)\psi_{\downarrow}(\vec r),
\end{eqnarray}
where $\hat{\xi}\equiv-\frac{\nabla^2}{2m}-\mu$, and $U(\vec r)>0$
is the attractive coupling constant between electrons, and $H_{imp}$
includes scattering with nonmagnetic impurities. When the pairing
interaction, $U(\vec r)$, is nonuniform, so is the order parameter in
this system. A standard technique to tackle this non-uniform
superconductivity problem is the quasiclassical Green's functions
\cite{Usadel,Kopnin,Schon1999}. In the dirty limit $\ell\ll\xi_0\sim\frac{\hbar v_F}{T_c}$, the quasiclassical Green's
functions obey a simple form of the Usadel equation, which in the
absence of a phase gradient is:
\begin{equation}\label{7}
\frac
D2\left(-\nabla^2\theta\right)=\Delta\cos\theta-\omega_n\sin\theta,
\end{equation}
where $D=\frac1d v_Fl$ is the diffusion constant, $l$ is the mean
free path, $d$ is the spatial dimension, and $\Delta$ is the
superconducting order parameter. $\theta$ is a real function of space
and Matsubara frequencies $\omega_n$ and is a parametrization of the
quasiclassical Green functions $g$ and $f$:
\begin{equation}\label{5}
g=\cos\theta,f=f^{\dag}=-i\sin\theta.
\end{equation}
Also, we list the relation between the integrated quasiclassical
Green's function and Gor'kov's Green's function $G$ and $F$:
\begin{equation}
g(\vec r)=\int\frac{d\Omega_p}{4\pi}\int\frac{d\xi_p}{i\pi}G(\vec
r,\vec p)=\frac1{i\pi N_F}\int\frac{d^3p}{(2\pi)^3}G(\vec r,\vec
p),\nonumber
\end{equation}
\begin{equation}\label{6}
f(\vec r)=\int\frac{d\Omega_p}{4\pi}\int\frac{d\xi_p}{i\pi}F(\vec r
,\vec p)=\frac1{i\pi N_F}\int\frac{d^3p}{(2\pi)^3}F(\vec r,\vec
p),\nonumber
\end{equation}
where $\vec r$ is the center of mass coordinate, and $\vec p$ is
momentum corresponding to the relative coordinate; $\Omega_p$ is the
angle of momentum $\vec p$ and $N_F$ is the density of states (per
spin) of the normal state at the Fermi energy. The
self-consistency equation reads:
\begin{equation}\label{8}
\Delta(\vec r)=U(\vec r)N_F\pi T\sum_n if_{\omega_n}(\vec r).
\end{equation}
For simplicity we assume the pairing is as given in Eq. (\ref{eq0}),
\[
U(\vec r)=\bar U+U_Q\cos(Qx).
\]

\subsection{The uniform pairing case}

Before analyzing the inhomogeneous pairing problem, let us briefly
review the calculation of $T_c$, the superconducting order parameter $\Delta(T=0)$, and
the DOS $\nu(E)$ of a dirty superconductor with a spatially uniform
coupling constant $U$, using quasiclassical Green's functions. In
this case Eqs. (\ref 7) and (\ref 8) admit a uniform solution
for both $\theta$ and $\Delta$:
\begin{equation}
\label{a1}
\theta=\arctan\left(\frac{\Delta}{\omega_n}\right).
\end{equation}
Using (\ref 8), we obtain the standard BCS
self-consistency equation:
\begin{equation}\label{a2}
1=UN_F\pi T\sum_n \frac{1}{\sqrt{\Delta^2+\omega_n^2}}.
\end{equation}\label{a3}
$T_c$ and $\Delta(T=0)$ are easily obtained from (\ref{a2}):
\begin{equation}
T_c=\frac{2C}{\pi}\omega_De^{-\frac{1}{UN_F}},\Delta_{(T=0)}=2\omega_De^{-\frac{1}{UN_F}}.\nonumber
\end{equation}
where $C=e^{\gamma}\approx 1.78$, with $\gamma=0.5772\ldots$  the
Euler constant, and $\omega_D$  the Debye frequency. The DOS can
be obtained from the retarded quasiclassical Green's function:
$\nu(E)=\textrm{Re}\{ g^R(E)\}$, which can be obtained from
$g(\omega_n)=\cos(\theta_n)$ by analytical continuation
$i\omega\rightarrow E+i0^+$:
\begin{equation}
\nu(E)=\textrm{Re}\frac{-iE}{\sqrt{\Delta^2-(E+i0^+)^2}}=\left\{\begin{array}{ll}\frac{E}{\sqrt{E^2-\Delta^2}},
& \textrm{if $E>\Delta$}\\0, & \textrm{if
$E<\Delta$}\end{array}\right..\nonumber
\end{equation}
Thus there exists a gap in the excitation spectrum $E_g=\Delta$, and
its ratio with $T_c$ is a universal number $\pi/C\approx1.76$. As
expected, these results for dirty superconductors are exactly the
same as those of clean superconductors, thus explicitly illustrating
Anderson theorem.

\section{The case of inhomogeneous pairing}\label{inhomo}

Using the formalism reviewed in the previous section, we now discuss
the non-uniform superconducting film. Our discussion will concentrate
on the limits of fast and slow pairing modulations, i.e., large and
small $Q\xi$ respectively ($\xi$ is the zero temperature coherence length in the dirty limit:
$\xi=\sqrt{{\hbar D}/{\bar{\Delta}_{T=0}}}\sim\sqrt{{\hbar D}/{T_c}}$, where $\bar{\Delta}$ is the spatially averaged $\Delta(x)$).

\subsection{Fast pairing modulation: proximity enhanced
  superconductivity \label{fast}}

With a nonuniform coupling $U(x)$, uniform
solution of either $\theta(x)$ or $\Delta(x)$ no longer exists. When fast
pairing modulation are present, the angle $\theta$ is dominated by its
$k=0$ Fourier component,
$\theta_0$, since it can not respond faster than its characteristic length scale
$\xi$. Corrections to the uniform solution are of the form $\theta_1\cos(Qx)$,
and are suppressed  by powers of $\frac{1}{Q\xi}$. From Eq.
(\ref 8), we see that in contrast to $\theta$, the order parameter
$\Delta(x)$ has
a factor of $U(x)$ in its
definition, and therefore it can fluctuate with the fast modulation of
$U(x)$. The modulating component of
$\Delta(x)$ is thus only suppressed by $U_Q/\bar U$, while the
modulating part of $\theta(x)$ is suppressed by both $U_Q/\bar U$
and $1/(Q\xi)$. Keeping both $1/Q\xi \ll 1$ and expanding in $U_Q/\bar U$, we can perturbatively solve Eqs. (\ref
7) and (\ref 8). Starting with:
\begin{equation}\label{9}
\Delta(x)=\Delta_0+\Delta_1\cos(Qx),\theta(x)=\theta_0+\theta_1\cos(Qx);
\end{equation}
Eq. (\ref 7) can be solved order by order:
\begin{eqnarray}\label{10}
\theta_0&=&\arctan\left(\frac{\Delta_0}{\omega_n}\right),\\
\theta_1
&=&\Delta_1\frac{\omega_n}{\frac
D2Q^2\sqrt{\omega_n^2+\Delta_0^2}+\omega_n^2+\Delta_0^2}.\nonumber
\end{eqnarray}
The self-consistency equation (\ref 8) can be Fourier transformed:
\begin{eqnarray}\label{15}
\Delta_0&=&N_F\pi T\sum_{\omega_n}\left(\bar
U\sin\theta_0+2\frac{U_Q}2\frac{\cos\theta_0}2\theta_1\right),\\
\frac{\Delta_1}2&=&N_F\pi T\sum_{\omega_n}\left({\bar
U}\frac{\cos\theta_0}2\theta_1+\frac{U_Q}2\sin\theta_0\right),\nonumber
\end{eqnarray}
where the $\omega_n$ index of $\theta_0$ and $\theta_1$ is implicit.

When $T\rightarrow T_c$, we can linearize $\theta_0$ and $\theta_1$
with respect to $\Delta_0$ and $\Delta_1$, respectively:
\begin{eqnarray}
\label{17}
\sin\theta_0\approx\frac{\Delta_0}{|\omega_n|},\hspace{3mm}
\theta_1(\cos\theta_0)\approx\frac{\Delta_1}{|\omega_n|+\frac{DQ^2}2}.\nonumber
\end{eqnarray}
Note that
\begin{equation}\label{18}
\sum_{n=0}^{N_0}\frac1{n+1/2}\approx\ln N_0+2\ln2+\gamma \textrm{ for
$N_0\gg1$,}
\end{equation}
where $\gamma$ is the Euler constant, we have approximately
\begin{eqnarray}
\label{19}
2\pi T\sum_{\omega_n=0}^{\frac{\omega_D}{2\pi
T}}\frac1{\omega_n}&\approx&\ln(2C\omega_D/\pi T),\\
2\pi T\sum_{\omega_n=0}^{\frac{\omega_D}{2\pi
T}}\frac1{\omega_n+DQ^2/2}&\approx&\ln\left(1+\frac{\omega_D}{DQ^2/2}\right),\nonumber
\end{eqnarray}
where, as before, $C=e^{\gamma}\approx1.78$ and $\omega_D$ is the Debye frequency.
Defining
\begin{equation}
\label{20}
K_0=\bar UN_F\ln(2C\omega_D/\pi T),\hspace{3mm}K_1=\bar
UN_F\ln\left(1+\frac{2\omega_D}{DQ^2}\right),
\end{equation}
we get
\begin{eqnarray}
\label{21}
\Delta_0&=&K_0\Delta_0+\frac12\frac{U_Q}{\bar U}K_1\Delta_1,\nonumber\\
\Delta_1&=&\frac{U_Q}{\bar U}K_0\Delta_0+K_1\Delta_1.\nonumber
\end{eqnarray}
$T_c$ is the temperature at which this equation admits a nonzero solution:
\begin{equation}\label{23}
T_c=\frac{2C}{\pi}\omega_D\exp\left(-\frac{1}{U_{eff}N_F}\right),
\end{equation}
where the effective pairing strength is:
\begin{equation}\label{24}
U_{eff}=\bar U\left(1+\left(\frac{U_Q}{\bar
U}\right)^2\frac{K_1}{2(1-K_1)}\right).
\end{equation}
This is the dirty case analogue of the result obtained by
Ref. \onlinecite{Podolsky}.

Next we turn to the order parameter. At $T=0$ the sums in the
self-consistency equations (\ref{15}) become integrals, which can be performed (see also Appendix \ref{appA}):
\begin{eqnarray}\label{31}
\Delta_0=N_F\bar
U\Delta_0\ln\left(\frac{2\omega_D}{\Delta_0}\right)+\frac12\frac{U_Q}{\bar
U}K_1\Delta_1, \nonumber\\
 \frac{\Delta_1}2= \frac{K_1\Delta_1}2
+\frac{N_FU_Q}2\Delta_0\ln\left(\frac{2\omega_D}{\Delta_0}\right),
\end{eqnarray}
thus giving the solution
\begin{eqnarray}\label{33}
\Delta_{0(T=0)}&=&2\omega_D\exp\left(-\frac1{U_{eff}N_F}\right),\nonumber\\
\Delta_{1(T=0)}&=&\Delta_{0(T=0)}\frac{U_Q}{U_{eff}}\frac1{1-K_1}.\nonumber
\end{eqnarray}
with the same $U_{eff}$ defined in (\ref{24}). Noting that
$\Delta_0$ is the spatially averaged value of the order parameter
$\bar{\Delta}$, we arrive at the conclusion that in the
limit $Q\xi\gg1$, the ratio
\begin{equation}\label{36}
\frac{2\bar{\Delta}}{T_c}=\frac{2\Delta_{0(T=0)}}{T_c}=\frac{2\pi}{C}
\end{equation}
is preserved.

The modification of the gap, however, must be addressed
separately. Although the gap and the order parameter coincide for
a uniform BCS superconductor, this is not generally true in a
nonuniform superconductor. To obtain the DOS and the gap one has
to rephrase the problem in a real-time formalism and calculate the
retarded Green's function which is parameterized by a complex
$\theta(x,E)=\theta'(x,E)+i\theta''(x,E)$ with both
$\theta',\,\theta''$ real, and then compute the DOS via
$\nu(x,E)=\textrm{Re}g^R(x,E)=\textrm{Re}\cos\theta(x,E)=\cos\theta'\cosh\theta''$\cite{Kopnin,Schon1999}.
Naively one can perform the prescription $i\omega\rightarrow
E+i0^+$ in the imaginary time Green's functions to obtain the
retarded ones, but our perturbative solution will break down as
$E$ approaches $\Delta_0$, since $\theta_1$ diverges faster than
$\theta_0$. Therefore to analyze the gap one has to re-solve the
real time counterpart of equation (\ref{7}) with $\Delta(x)$ given
above. Note that our solution of $\Delta(x)$ is still valid,
sparing us the need to solve the self-consistency equation.

\begin{figure}
\includegraphics[scale=0.5]{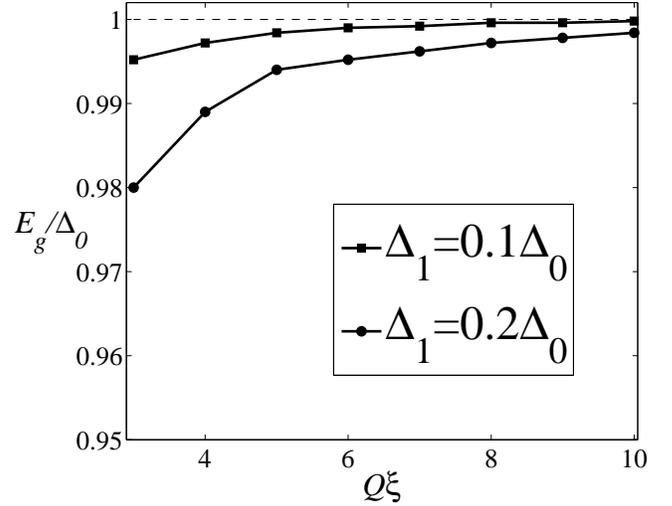}
     \caption{The energy gap, $E_g$, measured in units of $\Delta_0$, vs. $Q\xi$ for $Q\xi\gg1$.
     The two curves are for $\Delta_1/\Delta_0=0.1$ and $0.2$, respectively. Here, $\Delta_0$ and $\Delta_1$ are the uniform and oscillating components of the order parameter, respectively. $Q$ is the modulating wavevector of the inhomogeneous coupling constant;
        $\xi$ is the superconducting coherence length. The estimated numerical error of $E_g/\Delta_0$ is about $0.01$. The deviation of $E_g$ from $\Delta_0$ is small, but it increases with larger $\Delta_1/\Delta_0$ or smaller $Q\xi$.
        }\label{LargeQ}
\end{figure}

In real time, Eq. (\ref 7) becomes:
\begin{eqnarray}\label{num}
-\frac
D2\partial_x^2\theta'&=&\cos\theta'(\Delta\cosh\theta''-E\sinh\theta''),\nonumber\\
\frac
D2\partial_x^2\theta''&=&\sin\theta'(\Delta\sinh\theta''-E\cosh\theta'').
\end{eqnarray}
We numerically solved these coupled equations with periodic boundary
condition on $[0,2\pi/Q]$, and computed the DOS
$\nu(E)=\cos\theta_1\cosh\theta_2$, and thereby obtained the gap. We
find that despite the fluctuating $\Delta(x)$, the energy gap,
$E_g$, is spatially uniform. Fig. \ref{LargeQ} shows a graph of
$E_g$ vs. $Q\xi$ for $\Delta_1/\Delta_0=0.1$ and $0.2$. Again, in the plot we define the coherence length $\xi$ to be
$\sqrt{\hbar D/\bar{\Delta}_{T=0}}=\sqrt{\hbar D/{\Delta_{0,T=0}}}$. One can see
that in the limit $Q\xi\rightarrow\infty$ $E_g$ coincides with
$\Delta_0$, and nonzero $1/(Q\xi)$ brings about only small
corrections to make the gap slightly smaller than $\Delta_0$. These
corrections increase with smaller $Q\xi$ or larger $U_Q/\bar U$
(i.e., $\Delta_1/\Delta_0$). Thus we find that for $Q\xi\gg 1$ case
\begin{equation}
\frac{2E_{g(T=0)}}{T_c}\lesssim\frac{2\Delta_{0(T=0)}}{T_c}=\frac{2\pi}{C}=3.52.
\end{equation}
It is easy to understand the uniformity of $E_g$, since the wave
function of a quasiparticle excitation should be extended on a
length scale $1/Q\ll\xi$. Some intuition for the fact that
$E_g\approx\Delta_0$ is provided in Sec. \ref{sec3}.

\subsection{Slow pairing fluctuations: WKB-like local
  superconductivity}

When the pairing strength fluctuates slowly, i.e., over a large
distance, both the Green's functions and the order parameter $\Delta(x)$ can vary
on the length scale of $1/Q$, and we can approximate the zeroth
order solution by a 'local solution':
\begin{equation}
\label{37}
\theta_0(x)=\arctan\left(\frac{\Delta(x)}{\omega_n}\right),
\end{equation}
where $\Delta(x)$ is to be solved from the self-consistency
equation. This 'local' property of the system implies a large
spatial variation of both $\Delta(x)$ and $\theta(x)$, in contrast
to the $Q\xi\gg1$ case. To improve the zeroth order solution, we
write $\theta(x)=\theta_0(x)+\theta_1(x)$. Neglecting the small
gradient term of $\theta_1$, one can solve for $\theta_1$ from
Usadel's equation (\ref{7}) :
\begin{equation}\label{39}
\theta_1=\frac{D}2\left(\frac{\omega_n\partial_x^2\Delta}{(\Delta^2+\omega_n^2)^{3/2}}-\frac{2\Delta\omega_n(\partial_x\Delta)^2}{(\Delta^2+\omega_n^2)^{5/2}}\right),
\end{equation}
thus the self-consistency equation (\ref 8) becomes
\begin{equation}\label{40}
\Delta(x)=U(x)N_F2\pi T\sum_{n=0}^{\frac{\omega_D}{2\pi
T}}\left(\frac{\Delta}{\sqrt{\Delta^2+\omega_n^2}}+\frac{\omega_n}{\sqrt{\Delta^2+\omega_n^2}}\theta_1\right).
\end{equation}
In the Ginzburg-Landau regime, one is justified in keeping lowest
order terms in (\ref{40}):
\begin{eqnarray}\label{42}
\Delta(x)=U(x)N_F\left\{\Delta(x)\ln\left(\frac{2C\omega_D}{\pi
T}
\right)-\frac{7\zeta(3)}{8\pi^2T^2}\Delta^3(x)\right.\nonumber\\\left.+\frac{\pi\hbar
D}{8T}\partial_x^2\Delta(x) \right\},
\end{eqnarray}
where $\zeta(n)$ is the Riemann $\zeta$ function. Remarkably, equation
(\ref{42}) is nothing but the Ginzburg-Landau equation for a
modulating coupling constant $U(x)$ with $Q\xi\ll 1$, and is
precisely the dirty case analogue of equation (9) in
Ref. \onlinecite{Podolsky}, with $\xi$ replaced by the dirty limit
expression ${\tilde{\xi}}^2=\hbar\pi D/8T$
($\tilde{\xi}$ is slightly different from the coherence length defined in this work
$\xi\equiv\sqrt{\hbar D/\bar{\Delta}_{T=0}}$, where $\bar{\Delta}$ is the spatially averaged $\Delta(x)$). In the limit $Q\xi\rightarrow 0$,
$\Delta(x)$ would be determined only by the local value of $U(x)$,
and the mean field transition temperature would be given by
$T_{c,max}=2C\omega_D/\pi\exp(-1/(\bar U+U_Q))$. A small but
nonzero $Q\xi$ leads to a weak coupling between spatial regions,
hence slightly reducing the mean field $T_c$. Following the analysis
of Ref. \onlinecite{Podolsky}, one obtains the mean
field transition temperature:
\begin{equation}
\label{43}
T_c^{MF}\approx\frac{2C\omega_D}{\pi}e^{-{1}/{N_F(\bar U+U_Q
)}}e^{-\tilde{\xi} QA/\sqrt2},
\end{equation}
where $A\equiv\sqrt{U_Q/(N_F\bar U^2)}$.

\begin{figure}
    \centering
    \subfigure[]{\includegraphics[scale=0.4]{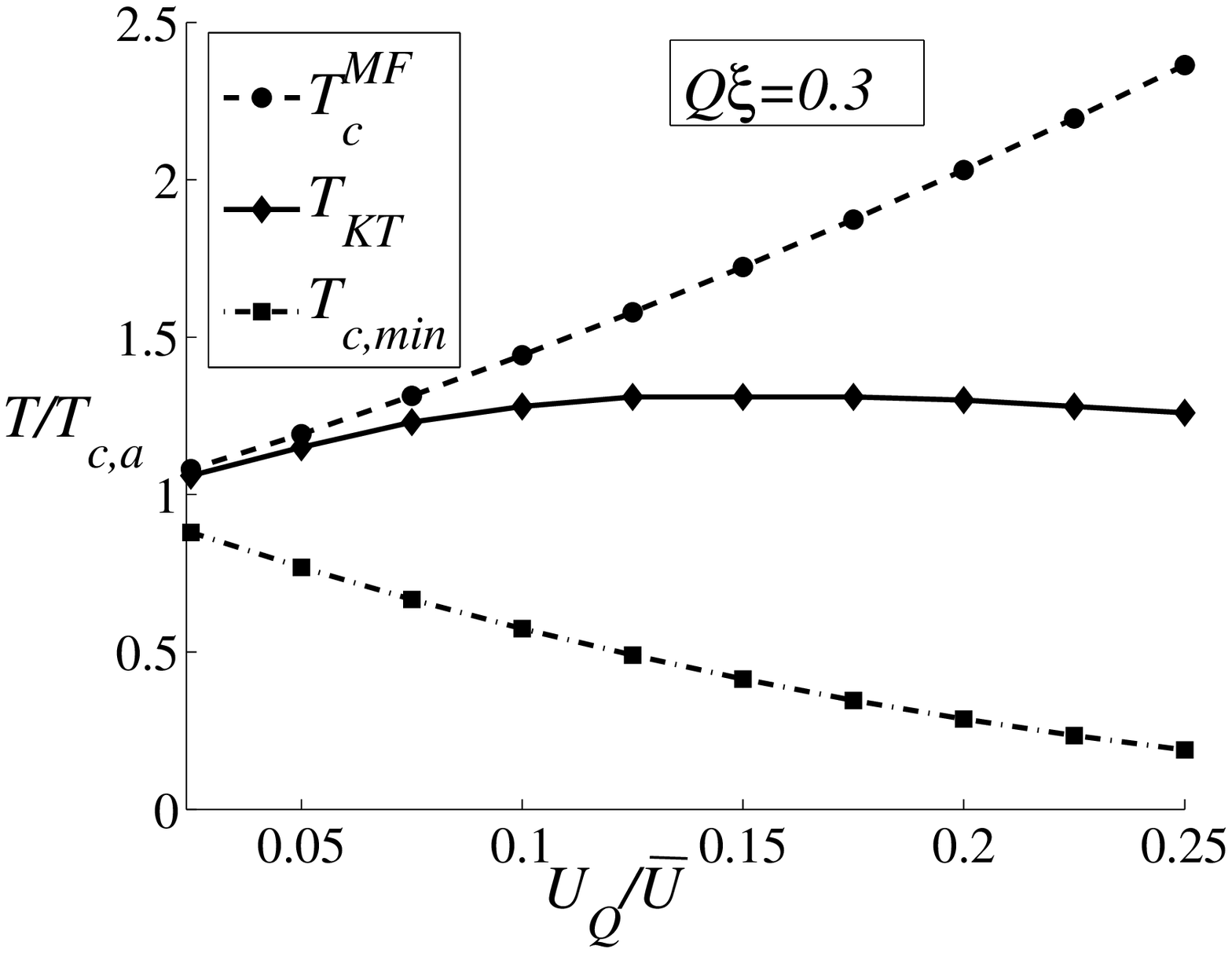}}
    \subfigure[]{\includegraphics[scale=0.4]{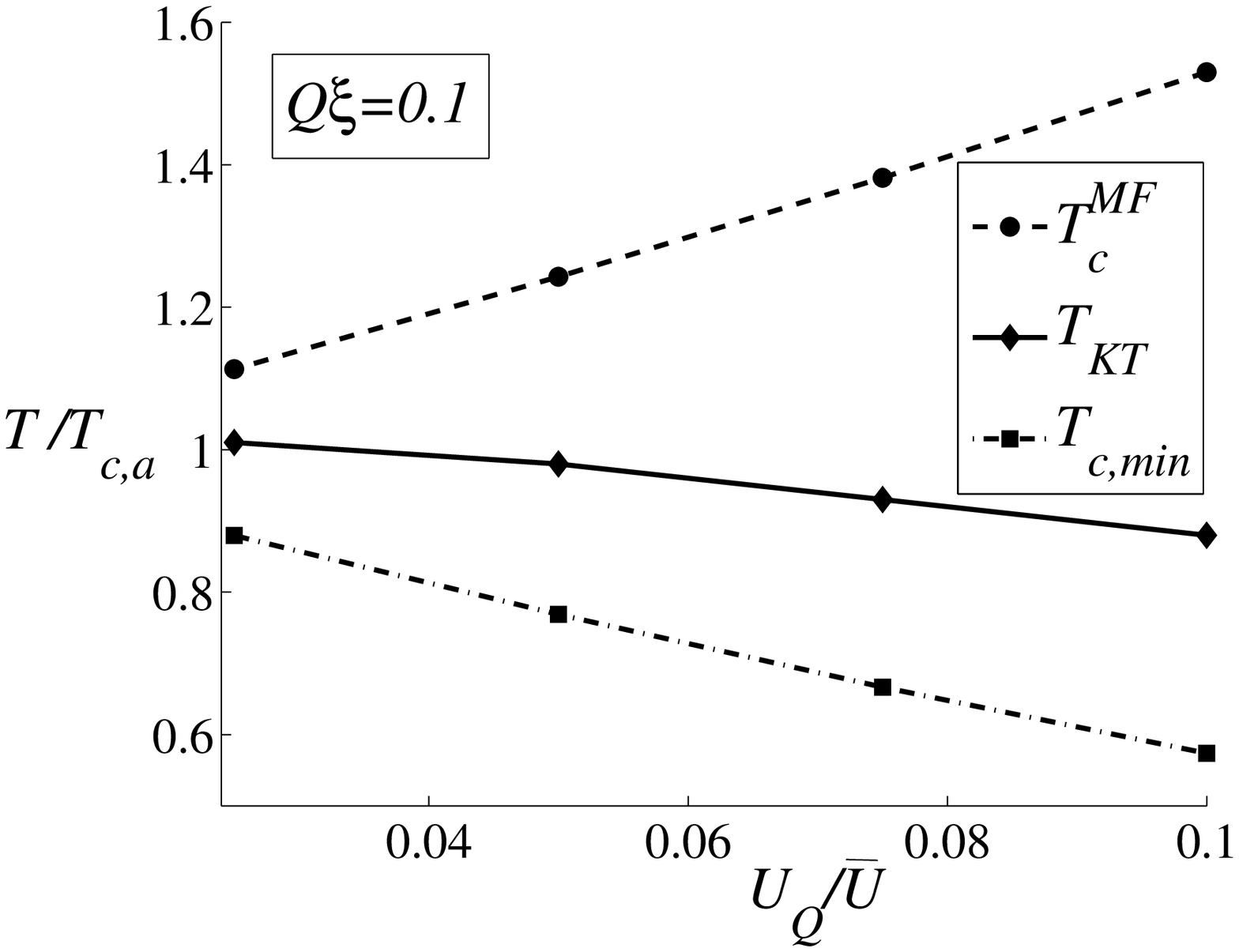}}
    \subfigure[]{\includegraphics[scale=0.4]{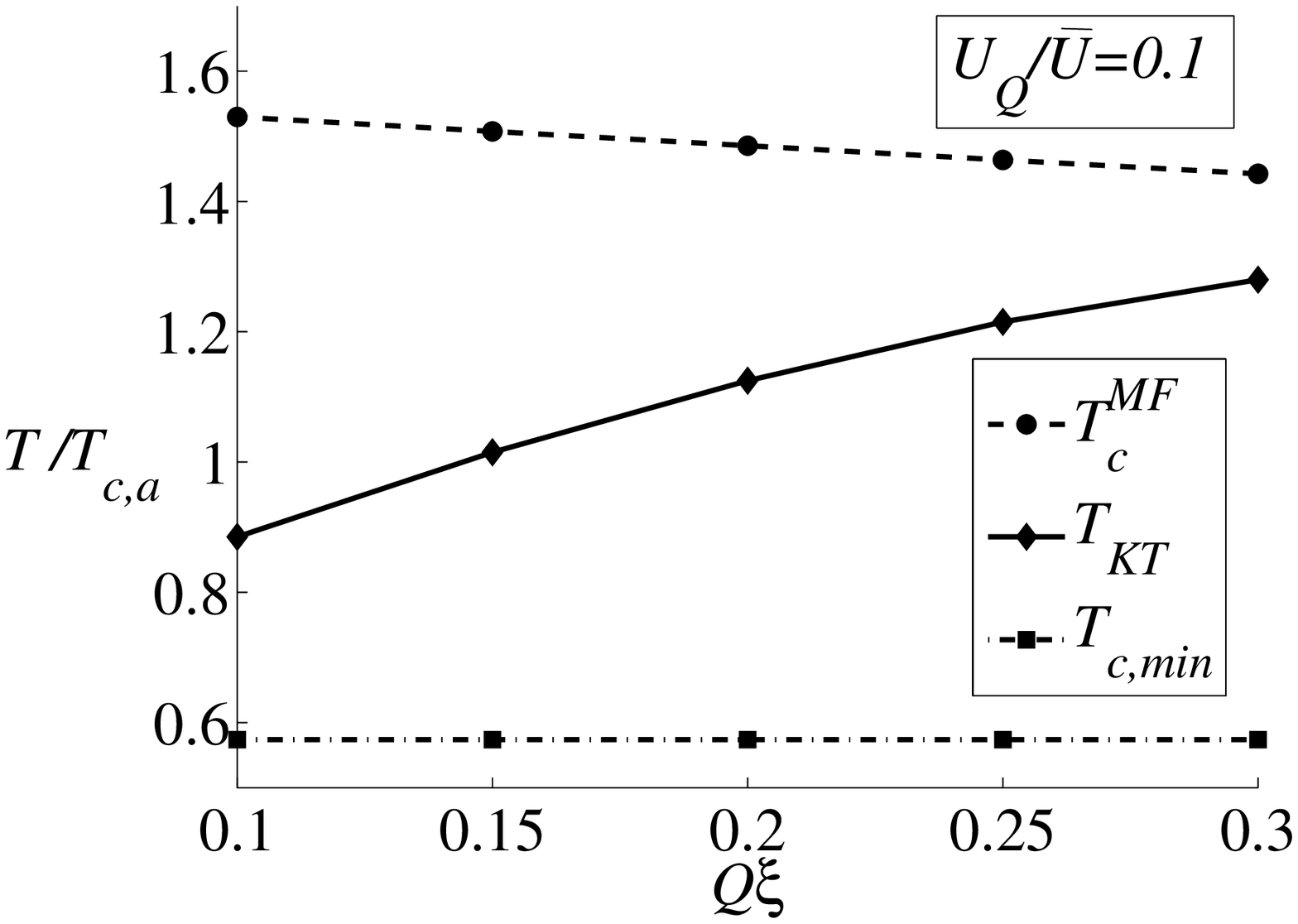}}
    \caption{The mean field transition temperature $T_c^{MF}$, the Kosterlitz-Thouless temperature $T_{KT}$, and the minimum mean field transition temperature $T_{c,min}$ (see below) (a) vs. $U_Q/\bar{U}$ with $Q\xi=0.3$;
    (b) vs. $U_Q/\bar U$ with $Q\xi=0.1$; (c) vs. $Q\xi$ with $U_Q/\bar U=0.1$. In
    all cases $\bar UN_F=0.2$. $T$ is in units of $T_{c,a}\equiv\frac{2C}{\pi}\omega_De^{-1/N_F\bar U}$. Dashed curve is $T_c^{MF}$ determined from equation (\ref{43}); dash-dotted curve is the minimum mean field $T_c$ given by $T_{c,min}=\frac{2C}{\pi}\omega_De^{-1/N_F(\bar U-|U_Q|)}$; solid curve is $T_{KT}$ obtained from numerically minimizing (\ref{200}) and then solving (\ref{201}). Here, $\bar U$ and $U_Q$ are the uniform and oscillating components of the coupling constant, respectively. $N_F$ is the density of states of the normal state; $Q$ is the modulating wavevector of the inhomogeneous coupling constant; $\xi$ is the superconducting coherence length. The estimated numerical error of $T_{KT}/T_{c,a}$ is about 0.01.}\label{TKT}
\end{figure}

Although the inhomogeneous
$U(x)$ largely increases the mean field $T_c$, it also makes the
system more susceptible to phase fluctuations. This effect will be
more pronounced in a two-dimensional superconductor, which we will
focus on now. A film becomes superconducting through a
Kosterlitz-Thouless transition. To determine the Kosterlitz-Thouless transition temperature,
$T_{KT}$, we note that the Ginzburg-Landau free energy corresponding
to (\ref{42}) is
\begin{eqnarray}
\label{200}
F(\Delta(x))&=&N_F\int d^3x\{\alpha(x)\Delta^2(x)+\frac{\beta}2\Delta^4(x)\nonumber\\&+&\gamma(\partial_x\Delta)^2)\},\\
\alpha(x)&=&\frac1{N_FU(x)}-\ln\left(\frac{2\times 1.78\omega_D}{\pi
T}\right),\nonumber\\\beta
&=&\frac{7\zeta(3)}{8\pi^2T^2},\gamma=\frac{\pi\hbar D}{8T}\nonumber.
\end{eqnarray}

As a functional of $\Delta(x)$, $F$ can be minimized numerically,
thus giving a solution of $\Delta(x)$. The free energy cost for
phase fluctuations is approximately $F=\frac12\int
d^2xJ(x)(\nabla\theta)^2$. For quasi-2d films,
\begin{equation}\label{stiffness}
J(x)=2N_{\bot}N_F^{2d}{\tilde{\xi}}^2|\Delta_{MF}(x)|^2,
\end{equation}
where $N_F^{2d}$ is the 2d electron DOS, $N_{\bot}$ is the number of channels,
$\tilde{\xi}\equiv\sqrt{\frac{\pi\hbar D}{8T}}$, and $\Delta_{MF}$ is the mean field solution
of (\ref{200}). To explain the bilayer thin film experiments investigated by Long et al.\cite{subgap,superweak},
we use the measured value of the diffusion constant $D=5\times10^{-3}m^2s^{-1}$ (see Ref. \onlinecite{subgap}), and estimate $N_{\bot}= k_Fd/\pi\approx50$, where the film thickness $d\approx10\sim20$nm\cite{subgap,superweak}, and the Fermi wave vector $k_F\sim1{\AA}^{-1}$. As in Ref. \onlinecite{Podolsky}, one can estimate
$T_{KT}$ self-consistently from
\begin{equation}\label{201}
T_{KT}=\frac{\pi}2\sqrt{\overline{J(x)}(\overline{1/J(x)})^{-1}},
\end{equation}
since $\overline{J(x)}$ is the stiffness along the "stripes",
while $(\overline{1/J(x)})^{-1}$ perpendicular to the "stripes".
Although our estimation of $N_{\bot}$ is crude, the value of $T_{KT}$ is very insensitive to it. This is because $T_{KT}$ is solved self-consistently from (\ref{201}). If one attempts to use a larger $N_{\bot}$ in (\ref{stiffness}), the enhancement of $T_{KT}$ is limited by $J(x)$ which itself is suppressed as temperature increases. Typical solutions of $T_{KT}$ are shown in
FIG. \ref{TKT}. One can see that the phase fluctuation region,
i.e. the difference between $T_c^{MF}$ and $T_{KT}$, increases
with stronger inhomogeneity (FIG. \ref{TKT}(a) and (b)). Also for
longer wave length modulation $T_{KT}$ is reduced more strongly
(FIG. \ref{TKT}(c)). Heuristically, this is because for smaller
$Q\xi$ the superconducting stripes become farther apart, and
therefore it is more difficult for them to achieve phase
coherence.

Moving our focus to the zero-temperature order parameter and gap, we
note that at $T=0$ the integrals in equation (\ref{40}) can be done:
\begin{equation}\label{48}
\frac{\Delta(x)}{U(x)N_F}=\Delta(x)\ln\left(\frac{2\omega_D}{\Delta(x)}\right)+\frac{\pi
D\partial_x^2\Delta}{8\Delta(x)}-\frac{\pi
D(\partial_x\Delta)^2}{16\Delta^2(x)},\nonumber
\end{equation}
This can be approximately solved by:
\begin{eqnarray}\label{49}
\Delta(x)&\approx&\Delta_0(x)e^{-\eta(x)},\\
\Delta_0(x)&=&2\omega_De^{-\frac{1}{N_FU(x)}},\nonumber\\
\eta(x)&=&\frac{\pi
D}{8\Delta_0(x)}Q^2A^2\left(\cos(Qx)-\frac12A^2\sin^2(Qx)\right).\nonumber
\end{eqnarray}
Note that $\partial_x\Delta(x)\approx-A^2Q\sin(Qx)\Delta(x)$ [with
$A$ defined under Eq. (\ref{43})], for our WKB analysis to be
self-consistent, we need to require the that $A \lesssim\mathcal
O(1)$, thus $U_Q/\bar U$ needs to be small. Also, when this is
satisfied, $\eta(x)$ leads to a slight averaging between
$\Delta(x)$, which is a manifestation of proximity effect.

\begin{figure}
    \centering
    \subfigure[]{\includegraphics[scale=0.5]{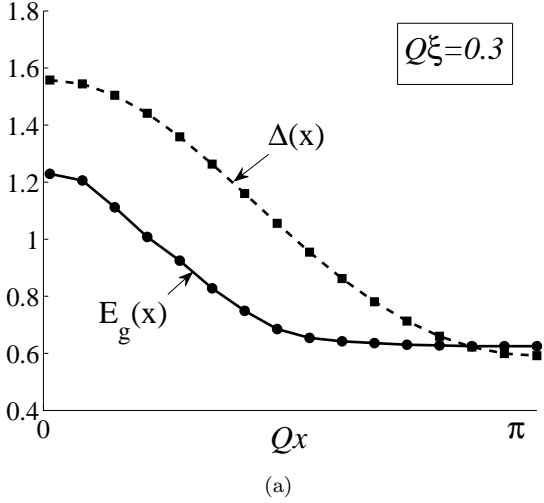}}
    \subfigure[]{\includegraphics[scale=0.5]{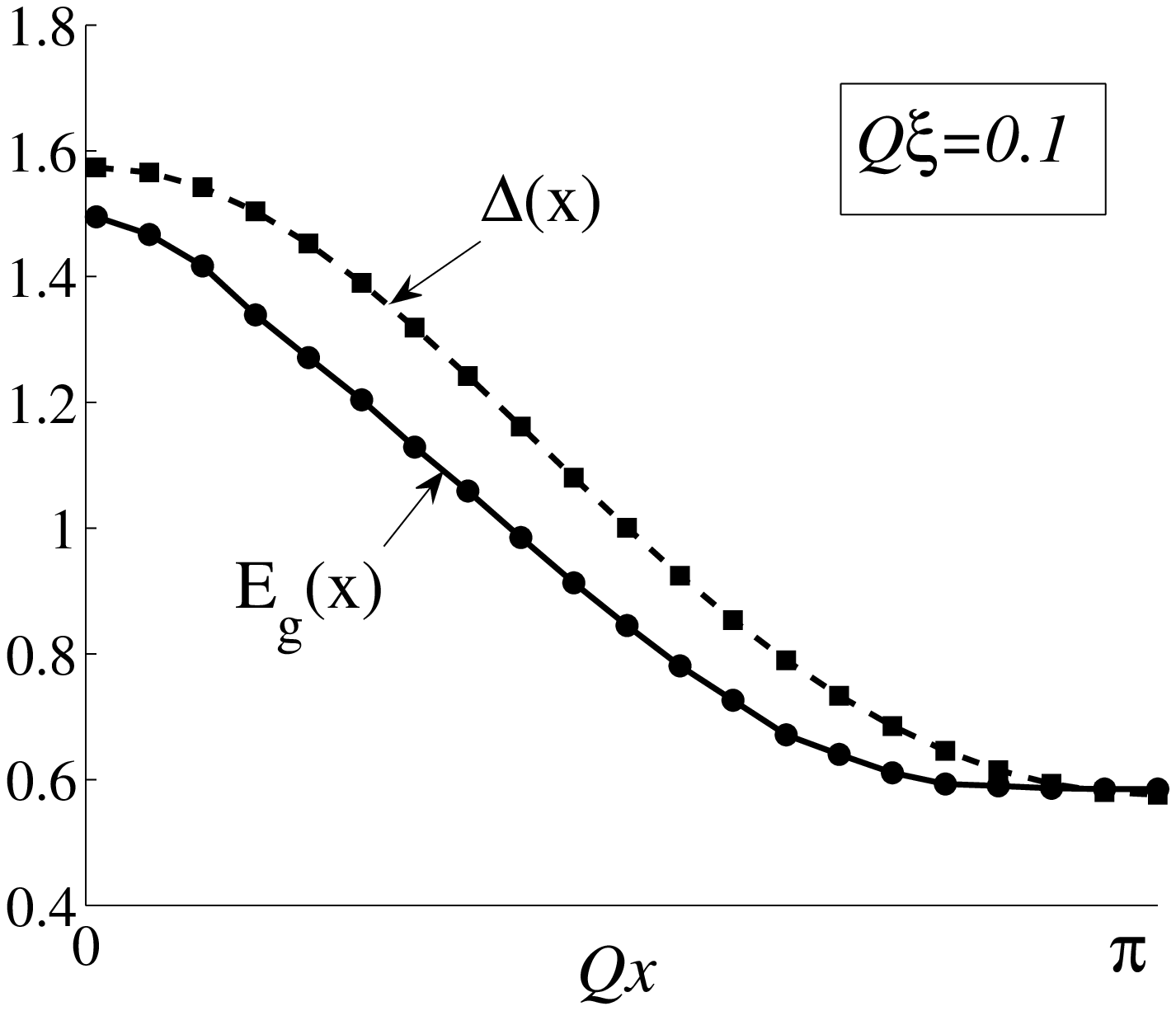}}
     \caption{The local order parameter $\Delta(x)$ and the local gap
$E_g(x)$ (in units of $\Delta(U_Q=0)=2\omega_De^{-1/\bar UN_F}$)
vs. spatial coordinate $x\in[0,\pi/Q]$. $Q\xi=0.3$ and $0.1$ in subfigure (a) and (b), respectively. $\bar UN_F=0.2$,
$U_QN_F=0.02$. Here, $\bar U$ and $U_Q$ are the uniform and oscillating components of the coupling constant, respectively. $N_F$ is the density of states of the normal state; $Q$ is the modulating wavevector of the inhomogeneous coupling constant; $\xi$ is the superconducting coherence length. The estimated numerical error of $E_g(x)$ is about $0.01$.}\label{gap}
\end{figure}

\begin{figure}
    \centering
    \subfigure[]{\includegraphics[scale=0.5]{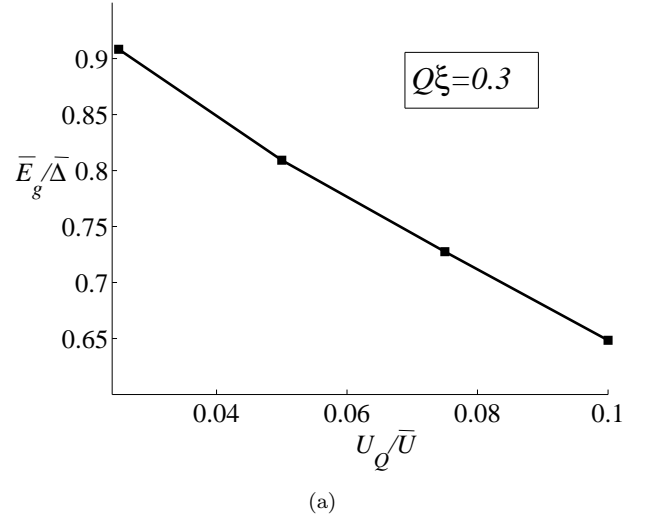}}
    \subfigure[]{\includegraphics[scale=0.5]{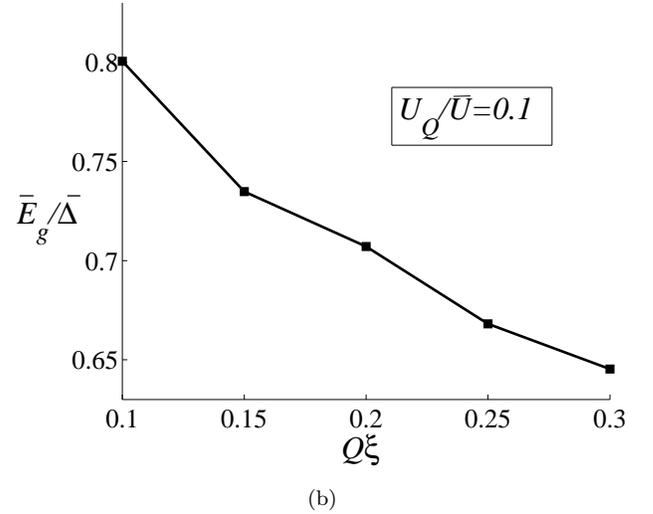}}
     \caption{The ratios of the spatially averaged gap
$\bar{E_g}$ to the spatially averaged $\bar{\Delta}$ (in units of
$2\omega_De^{-1/\bar UN_F}$) (a) vs. $U_Q/\bar U$ with
$Q\xi=0.3$; (b) vs. $Q\xi$ with $U_Q/\bar U=0.1$. $\bar UN_F=0.2$
in all cases. Here, $\bar U$ and $U_Q$ are the uniform and oscillating components of the coupling constant, respectively. $N_F$ is the density of states of the normal state; $Q$ is the modulating wavevector of the inhomogeneous coupling constant; $\xi$ is the superconducting coherence length. The estimated numerical error is about 0.01. }\label{gap_delta}
\end{figure}

\begin{figure}
    \centering
    \subfigure[]{\includegraphics[scale=0.5]{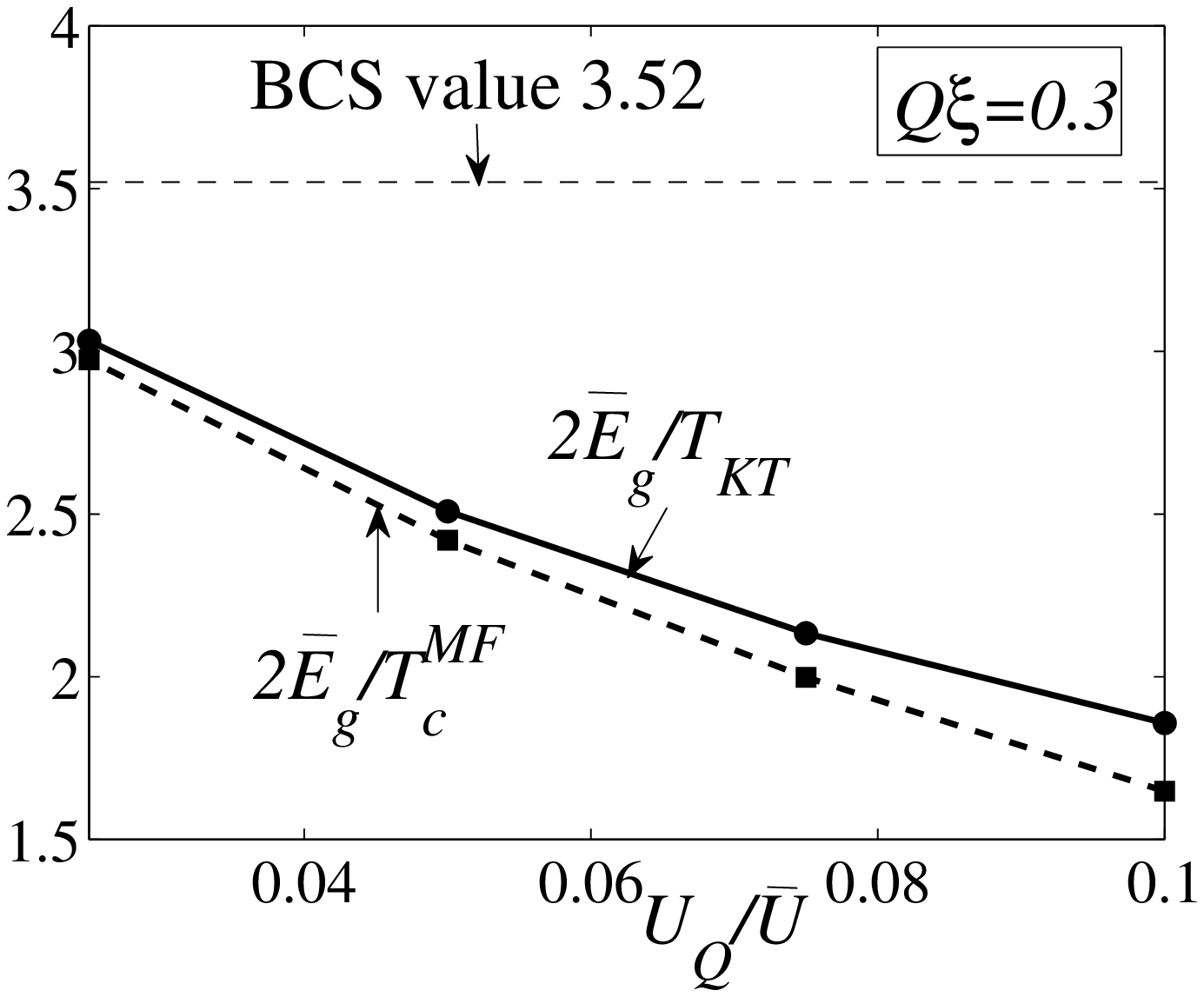}}
    \subfigure[]{\includegraphics[scale=0.5]{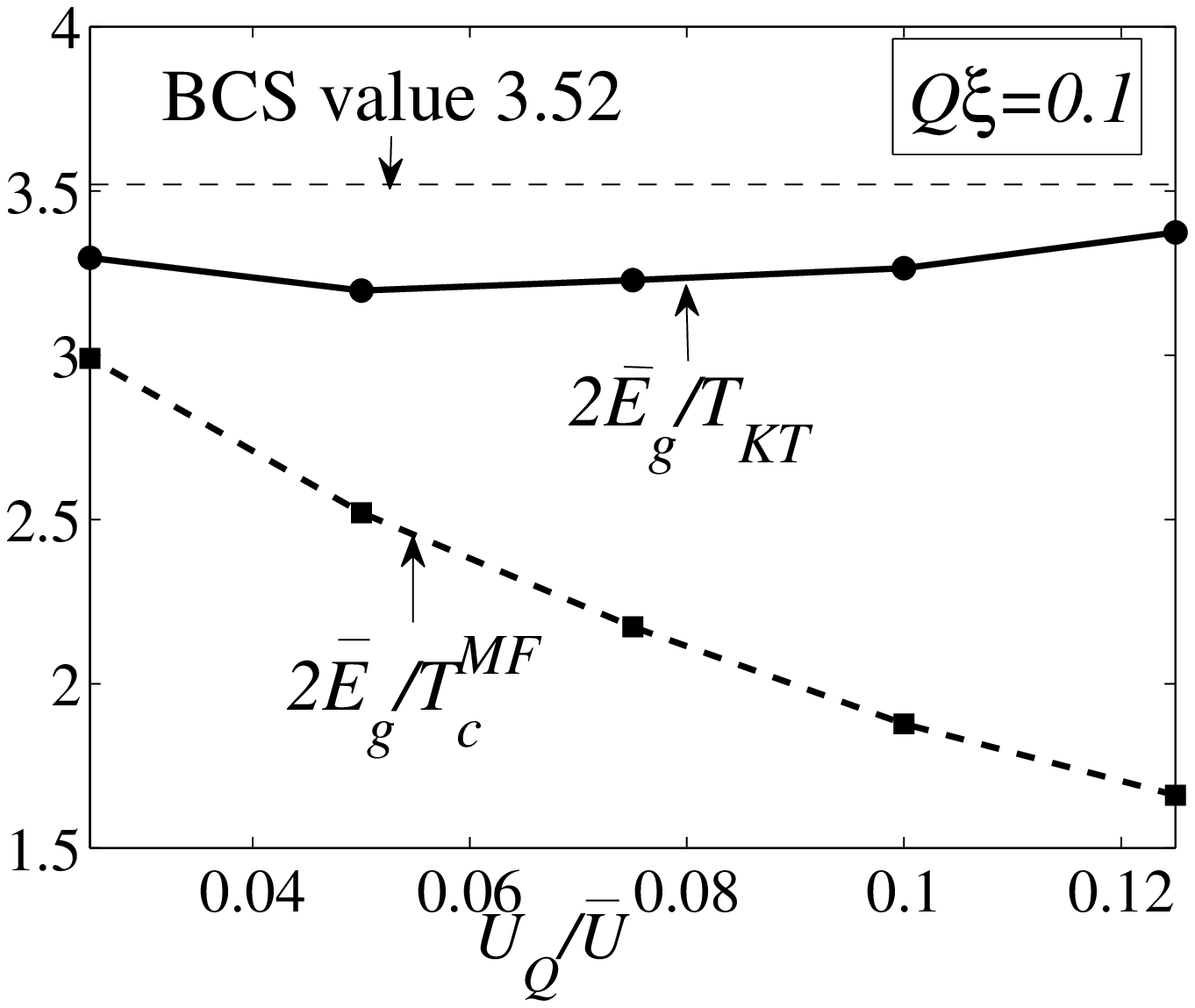}}
    \subfigure[]{\includegraphics[scale=0.5]{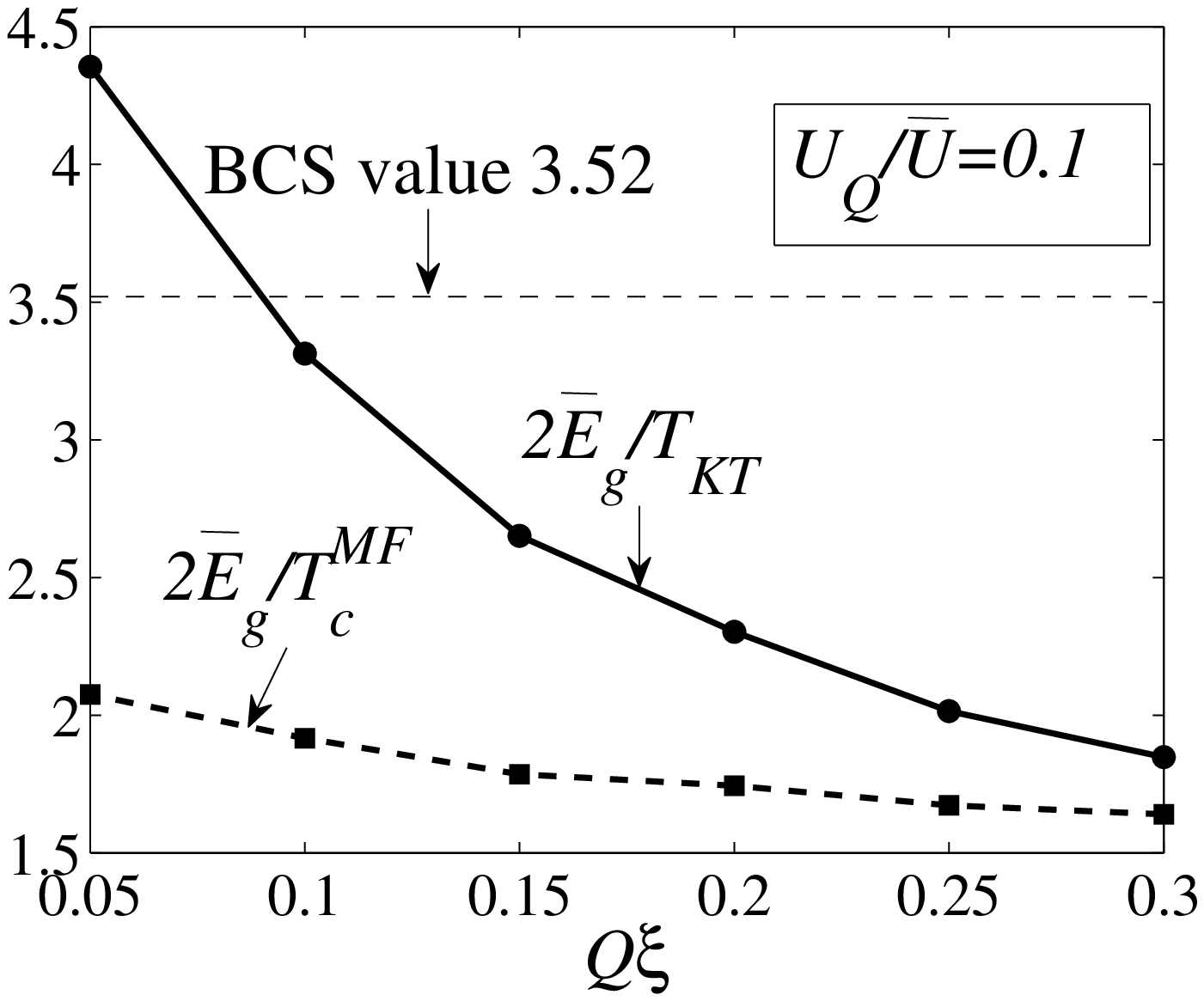}}
     \caption{The ratios of the spatially averaged gap $\bar{E_g}$  to
     $T_c^{MF}$ or $T_{KT}$ (a)vs. $U_Q/\bar{U}$, $Q\xi=0.3$; (b) vs. $U_Q/\bar U$, $Q\xi=0.1$;
     (c) vs. $Q\xi$, $U_Q/\bar U=0.1$. In all cases $\bar UN_F=0.2$. Here, $\bar U$ and $U_Q$ are the uniform and oscillating components of the coupling constant, respectively. $N_F$ is the density of states of the normal state; $Q$ is the modulating wavevector of the inhomogeneous coupling constant; $\xi$ is the superconducting coherence length. The estimated numerical error is about 0.02. }\label{ratio}
\end{figure}

To analyze the gap, we must switch to a real time formalism again,
since our perturbative solution for the Green's function becomes invalid as
$E\rightarrow\Delta(x)$. Thus we have to solve the real time Usadel
equation (\ref{num}) with $\Delta(x)$ obtained above. Using the same
numerical code as in Sec. \ref{fast}, we have obtained the local
gap $E_g(x)$, which is plotted vs. $x$ in FIG. \ref{gap} for half
a period of modulation. One can see that in general $E_g(x)$ is
lower than $\Delta(x)$, and when $Q\xi=0.3$, $E_g(x)$ is largely set
by the region with weakest coupling; but when $Q\xi\rightarrow0$,
$E_g(x)$ tends to follow much closer to $\Delta(x)$ as expected. In addition, the minimum of $E_g(x)$ is
always slightly higher than the minimum of $\Delta(x)$ by an amount
that also diminishes upon $Q\xi\rightarrow0$. This behavior will be
further clarified in the next section.

The ratio $\bar{E_g}/\bar{\Delta}$ vs. $U_Q/\bar U$ or $Q\xi$
is plotted in FIG. \ref{gap_delta}. The
suppression of the gap strengthens when either the inhomogeneity becomes
stronger ($U_Q/\bar U$ is large) or its length scale $L\sim1/Q$ becomes
smaller, consistent the results in FIG. \ref{gap}.
The $\bar{E_g}$ suppression relative to
$\bar{\Delta}$, together with the fact the $T_c^{MF}$ is largely determined by
strongest-coupling region, implies that the ratio
$2\bar{E_g}/T_c^{MF}$ is generically reduced. The ratios $2\bar{E_g}/T_c^{MF}$ and $2\bar{E_g}/T_{KT}$ are plotted in FIG. \ref{ratio} for several representative cases. As expected, there is always a
strong suppression of the ratio $2\bar{E_g}/T_c^{MF}$ from $3.52$;
for a two-dimensional system, however, the ratios with $T_{KT}$ are
more subtle: for very small $Q\xi$ the ratio $2\bar{E_g}/T_{KT}$
might be enhanced due to the large deviation of $T_{KT}$ from
$T_c^{MF}$ (see also FIG. \ref{TKT}(c)), while for larger value of
$Q\xi$ the phase fluctuation region is narrow(see also
FIG. \ref{TKT}(a)), and $2\bar{E_g}/T_{KT}$ is reduced from $3.52$.

For the purpose of comparison with the thin film experiments, a
comment on the determination of $T_c^{MF}$ and $T_{KT}$ is in
order. Due to disorder and phase fluctuations, the resistive
transition curve can be significantly broadened. $T_c^{MF}$ can be
estimated as the temperature at which the resistance drops to half
of its normal state value, while $T_{KT}$ can be defined as the
temperature at which the resistance drops below the measurement
threshold (see, for example, Ref. \onlinecite{Hsu1998}).
Alternatively, one can extract $T_c^{MF}$ from fitting the
fluctuation resistance to Aslamazov-Larkin theory\cite{AL}, and
obtain $T_{KT}$ from nonlinear I-V characteristics or from fitting
the resistance below $T_c^{MF}$ to Halperin-Nelson
form\cite{HaperinNelson} (see, e.g., Refs.
\onlinecite{Hebard1985,Mooij}). Thus both $T_c^{MF}$ and $T_{KT}$
in principle can be measured from experiments, and can be used for
comparison with our theoretical results here.

\subsection{Additional inhomogeneities}

Apart from modulation of the coupling $U$, one may also be interested
in a simultaneous modulation of other properties. For example, in
the small $Q\xi$ limit, one may expect the periodicity of $U$ to be
accompanied by a periodicity of the local density of states at the
fermi level, or the mean free path. Another possible modulation,
that of a periodic potential, is suggested in \cite{Podolsky}, and
in practice is equivalent to local modulation of $U$. Indeed, one
may use an effective description of the self consistency equation
\eqref{8}, taking $N_F\rightarrow N_F+N_Q\cos(Qx)$ to lowest order
in the amplitude $N_Q$ of the local DOS in the form:
\begin{equation}
\Delta(\vec r)={N_F}U_{mod}(\vec r)\pi T\sum_n if_{\omega_n}(\vec
r).
\end{equation}
where
$U_{mod}=\overline{U}+{N_Q\overline{U}+\overline{N_F}U_Q\over
\overline{N_F}}\cos(Qx)$, and $N_F$ is the spatially averaged DOS.
Formally this is exactly the same as Eq. (\ref{eq0}), and can be
treated similarly, taking
\begin{eqnarray}
  U_Q\rightarrow {N_Q\overline{U}+ {N_F}U_Q\over
 {N_F}}
\end{eqnarray}
In practice, a local periodic potential may be imposed on the
system externally by either acoustic means or an electromagnetic
field. Thus it might be interesting to check the change in $T_C$
of a superconductor in the presence of an acoustic wave
experimentally.

Another possibility of interest is that along with $U$ the electron
mean-free path is modulated in the system. This can be naturally
occurring if the periodicity in $U$ is a consequence of spatial
variation in the properties of the material used. Alternatively, one
may obtain this case by a periodic doping of the superconductor.

In this case we may describe the system effectively by modification
of the Usadel equation \eqref{7} to:
\begin{equation}
-\frac12 \nabla\cdot
(D\nabla\theta)=\Delta\cos\theta-\omega_n\sin\theta,
\end{equation}
and taking the diffusion coefficient $D$ to be spatially dependent.
Choosing $D=\overline{D}+D_Q\cos(Qx)$ and repeating the treatment
above, we find that $D_Q$ does not change the values of the Green's
functions $\theta_0, \theta_1$ above (It however appears at higher
orders of the equation), and so doesn't change the results of this
paper within this order.

\section{Superconductor-normal-metal (SN) superlattice analogy\label{sec3}}

Some insight into the nature of
the lowest-lying excitations for both large and small $Q\xi$ cases
can be gained by considering a simplified system:
superconductor-normal-metal-superconductor (SNS) junctions. First,
consider a single SNS junction with length $L=2\pi/Q$, and
$\Delta(x)=\Delta$, $0$ in the S, N part respectively. Andreev bound states
will form in the normal metal, and the energy of these states can be
obtained by solving Bogoliubov-de Gennes (BdG) equations for the
clean case, or Usadel equations for the dirty case. In the limit
$L\rightarrow0$, the energy of the lowest-lying state is $\Delta$,
while in the opposite limit $L\gg\xi$, the (mini)gap is much smaller
than $\Delta$: in the clean case $E_g\sim v_F/L\sim(Q\xi)\Delta$ and
in the dirty case the gap equals the Thouless energy
$D/L^2\sim(Q\xi)^2\Delta$ \cite{deGennes,Zagoskin,Zhou}. These
states exponentially decay into the superconductors for a distance
$\sim\xi$.

Based on a single SNS junction, one can build an SN superlattice with
alternating superconductor and normal metal, each with length
$L=2\pi/Q$, and $\Delta(x)=\Delta$, $0$ in the S, N part respectively. If
$L\gg\xi$, Andreev bound states remain localized in the normal
regions with the gap much smaller than $\Delta$. On the other hand
if $L\ll\xi$, these states strongly mix with each other, and they
form a tight-binding band. Therefore the gap, namely the lower band
edge, is lower than $\Delta$, and in the limit
$Q\xi\rightarrow\infty$ it is precisely at $\Delta/2$, the averaged
$\Delta(x)$ (see the analytical calculation by Ref. \onlinecite{sn}).
The SN
superlattice thus allows a qualitative understanding of the gap's
behavior in the problem we addressed above: if
$Q\xi\gg1$, all excitations are extended in space, with the uniform
gap $E_g\approx\bar{\Delta}$; if $Q\xi\ll1$, the lowest-lying
excitations are localized in the weakest coupling regions whose gap
is close to the minimum of $\Delta(x)$. This analogy also elucidates
the features in FIG. \ref{gap}: given a point in space $x_0$, $E_g(x_0)$ is generally
lower than $\Delta(x_0)$, because the wave function of the low-lying
excitations originating at a nearby region (within $\sim\xi$) with smaller
$\Delta(x)$ are exponentially suppressed at $x_0$, and when $\xi$ is
smaller this effect is reduced; thus $E_g(x)$ follows closer to $\Delta(x)$ in the limit $Q\xi\rightarrow0$.
Finally, the difference between the minimum of $E_g(x)$ and the minimum
of $\Delta(x)$ resembles the minigap in SN superlattice $\sim v_F/L$
or $D/L^2$, which approaches zero as $Q\xi\rightarrow0$.

\section{Summary and Discussion}\label{summary}

In this paper we investigated the properties of dirty BCS
superconductors with a fluctuating pairing coupling constant
$U(x)=\bar{U}+U_Q\cos(Qx)$. Particularly, we analyzed the change in
the mean field $T_c$, the zero-temperature order parameter
$\Delta(x)$, and the energy gap in quasiparticle excitation $E_g(x)$
using the Usadel equation for quasiclassical Green's functions. In
addition, we estimated the Kosterlitz-Thouless transition
temperature $T_{KT}$. Our analysis found four different
regimes:\newline (1) $Q\xi\rightarrow\infty$. In this case the mean
field $T_c$ and the spatially averaged order parameter
$\bar{\Delta}$ are determined by the effective coupling constant
$U_{eff}\gtrsim\bar{U}$ [see Eq. (\ref{24})]. Moreover, since in
this regime any quasiparticle wavefunction is extended over the
length scale $L=1/Q$, the local energy gap $E_g$ is uniform in
space, and we found it to coincide with the spatially averaged
$\bar{\Delta}$. The ratios $2\bar{\Delta}/T_c=2E_g/T_c=3.52$
maintain their universal BCS value.\newline (2) $Q\xi\gtrsim 1$. In
this regime the physics is qualitatively the same as that of the
previous case. The gap $E_g$, however, is smaller than
$\bar{\Delta}$ by an amount that grows with decreasing $Q\xi$ or
increasing $U_Q/\bar U$. Therefore $2\bar{E_g}/T_c\lesssim3.52$
(see FIG. \ref{LargeQ}).\newline (3) $Q\xi\lesssim1$. The system
tends to divide into regions which behave according to the the local
value of $U(x)$. Thus the mean field $T_c$ is determined by the
first formation of local superconductivity upon lowering
temperature, and therefore $T_c^{MF}$ is close to highest 'local
$T_c$'. In contrast, the global energy gap or the spatially averaged
local gap is largely determined by the region with smallest $U(x)$.
Consequently, in this regime the ratio
$2\bar{E_g}/T_c^{MF}$ is always suppressed from the universal BCS
value, 3.52 (see FIG. \ref{ratio}a). Moreover, although the system
is affected by phase fluctuations, in this regime $T_{KT}$ is close
to $T_c^{MF}$ for small values of $U_Q$ (see FIG. \ref{TKT}a). Thus
$2\bar{E_g}/T_{KT}$ is also smaller than
3.52 (see FIG. \ref{ratio}a).\newline (4) $Q\xi\rightarrow0$. As
opposed to the previous regime, here
  phase fluctuations lead to a large suppression of
$T_{KT}$ relative to $T_c^{MF}$ (see FIG. \ref{TKT}b). Although
$2\bar{E_g}/T_c^{MF}$ is still below 3.52, the ratios $2\bar{E_g}/T_{KT}$
is close to or larger than $3.52$ (see FIG. \ref{ratio}c).

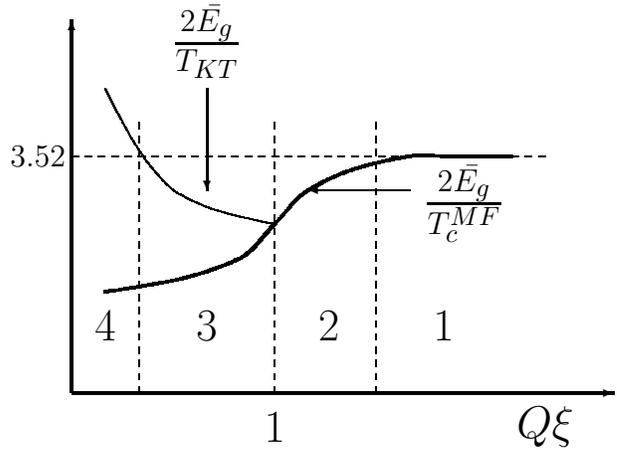
\begin{figure}
\ifx\JPicScale\undefined\def\JPicScale{0.9}\fi \unitlength
\JPicScale mm
\begin{picture}(85,65)(0,0)
\put(75,5){\makebox(0,0)[cc]{{\LARGE $Q\xi$}}}

\put(0,45){\makebox(0,0)[cc]{{\large 3.52}}}

\put(63,38){\makebox(0,0)[cc]{{\LARGE
$\frac{2\bar{E_g}}{T_c^{MF}}$}}}

\put(25,62){\makebox(0,0)[cc]{{\LARGE $\frac{2\bar{E_g}}{T_{KT}}$}}}

\linethickness{0.35mm} \put(5,10){\line(1,0){80}}
\put(85,10){\vector(1,0){0.12}} \linethickness{0.35mm}
\put(5,10){\line(0,1){55}} \put(5,65){\vector(0,1){0.12}}
\linethickness{0.2mm} \multiput(5,45)(1.97,0){36}{\line(1,0){0.99}}
\linethickness{0.3mm} \qbezier(70,45)(69.85,45)(68.15,45)
\qbezier(68.15,45)(66.44,45)(65,45)
\qbezier(65,45)(63.72,45)(62.5,45)
\qbezier(62.5,45)(61.28,45)(60,45)
\qbezier(60,45)(58.71,45.03)(57.49,45.1)
\qbezier(57.49,45.1)(56.27,45.18)(55,45)
\qbezier(55,45)(50.97,44.38)(47.22,43.25)
\qbezier(47.22,43.25)(43.47,42.11)(40,40)
\qbezier(40,40)(38.49,39.01)(37.37,37.67)
\qbezier(37.37,37.67)(36.25,36.32)(35,35)
\qbezier(35,35)(33.75,33.67)(32.65,32.3)
\qbezier(32.65,32.3)(31.54,30.94)(30,30)
\qbezier(30,30)(24.7,27.4)(17.69,26.19)
\qbezier(17.69,26.19)(10.67,24.99)(10,25)

\linethickness{0.1mm} \qbezier(35,35)(34.47,35)(29.12,36.26)
\qbezier(29.12,36.26)(23.77,37.53)(20,40)
\qbezier(20,40)(16.15,43.4)(13.16,48.91)
\qbezier(13.16,48.91)(10.16,54.41)(10,55)

\linethickness{0.2mm} \multiput(35,10)(0,1.95){21}{\line(0,1){0.98}}
\linethickness{0.2mm} \multiput(15,10)(0,1.95){21}{\line(0,1){0.98}}
\linethickness{0.2mm} \multiput(50,10)(0,1.95){21}{\line(0,1){0.98}}
\put(60,20){\makebox(0,0)[cc]{\LARGE 1}}

\put(43,20){\makebox(0,0)[cc]{\LARGE 2}}

\put(25,20){\makebox(0,0)[cc]{\LARGE 3}}

\put(10,20){\makebox(0,0)[cc]{\LARGE 4}}

\linethickness{0.2mm} \put(40,40){\line(1,0){15}}
\put(40,40){\vector(-1,0){0.12}} \linethickness{0.2mm}
\put(25,40){\line(0,1){15}} \put(25,40){\vector(0,-1){0.12}}
\put(35,5){\makebox(0,0)[cc]{\LARGE 1}}

\end{picture}
\caption{Schematic plot of the ratios $2\bar{E_g}/T_c^{MF}$ and
        $2\bar{E_g}/T_{KT}$ vs. $Q\xi$.
        Here $\bar{E_g}$ is the spatially averaged gap in local DOS; $T_c^{MF}$ is the mean field $T_c$;
        $T_{KT}$ is the Kosterlitz-Thouless transition temperature in 2d;
        $Q$ is the modulating wavevector of the inhomogeneous coupling constant;
        $\xi$ is the superconducting coherence length.
        1,2,3, and 4 are labels of different regimes described in the text.}\label{result}
\end{figure}

\begin{figure}

\ifx\JPicScale\undefined\def\JPicScale{0.8}\fi \unitlength
\JPicScale mm
\begin{picture}(100,80)(0,0)
\linethickness{0.3mm} \put(10,10){\line(1,0){90}}
\put(100,10){\vector(1,0){0.12}} \linethickness{0.3mm}
\put(10,10){\line(0,1){70}} \put(10,80){\vector(0,1){0.12}}
\linethickness{0.3mm} \multiput(10,60)(1.98,0){41}{\line(1,0){0.99}}
\linethickness{0.3mm} \multiput(10,30)(1.98,0){41}{\line(1,0){0.99}}
\linethickness{0.25mm} \qbezier(10,60)(10.15,60)(11.85,60)
\qbezier(11.85,60)(13.56,60)(15,60)
\qbezier(15,60)(16.28,60.02)(17.51,60.08)
\qbezier(17.51,60.08)(18.73,60.14)(20,60)
\qbezier(20,60)(25.3,59.44)(30.33,58.48)
\qbezier(30.33,58.48)(35.36,57.52)(40,55)
\qbezier(40,55)(41.57,54.09)(42.76,52.79)
\qbezier(42.76,52.79)(43.95,51.5)(45,50)
\qbezier(45,50)(46.54,47.58)(47.33,44.85)
\qbezier(47.33,44.85)(48.12,42.11)(50,40)
\qbezier(50,40)(52.06,38.02)(54.64,36.93)
\qbezier(54.64,36.93)(57.22,35.84)(60,35)
\qbezier(60,35)(68.42,32.56)(78.74,31.28)
\qbezier(78.74,31.28)(89.06,30.01)(90,30) \linethickness{0.2mm}
\qbezier(45,50)(44.84,50.04)(43.13,50.15)
\qbezier(43.13,50.15)(41.41,50.26)(40,50)
\qbezier(40,50)(37.21,49.32)(34.77,48.01)
\qbezier(34.77,48.01)(32.33,46.71)(30,45)
\qbezier(30,45)(28.52,43.94)(27.29,42.71)
\qbezier(27.29,42.71)(26.06,41.48)(25,40)
\qbezier(25,40)(23.42,37.62)(22.5,35)
\qbezier(22.5,35)(21.58,32.38)(20,30)
\qbezier(20,30)(18.75,28.39)(16.97,26.76)
\qbezier(16.97,26.76)(15.18,25.13)(15,25)

\put(45,5){\makebox(0,0)[cc]{\LARGE 1}}
\put(90,5){\makebox(0,0)[cc]{\LARGE $Q\xi$}}

\put(3,60){\makebox(0,0)[cc]{$T_{c,max}$}}

\put(5,30){\makebox(0,0)[cc]{$T_{c,a}$}}

\linethickness{0.15mm} \multiput(45,10)(0,2){28}{\line(0,1){1}}
\linethickness{0.15mm} \multiput(60,10)(0,2){28}{\line(0,1){1}}
\linethickness{0.15mm} \multiput(25,10)(0,2){28}{\line(0,1){1}}
\put(70,20){\makebox(0,0)[cc]{\LARGE 1}}

\put(53,20){\makebox(0,0)[cc]{\LARGE 2}}

\put(35,20){\makebox(0,0)[cc]{\LARGE 3}}

\put(17,20){\makebox(0,0)[cc]{\LARGE 4}}

\put(68,50){\makebox(0,0)[cc]{\large $T_c^{MF}$}}

\put(17,50){\makebox(0,0)[cc]{\large $T_{KT}$}}

\linethickness{0.3mm} \put(65,35){\line(0,1){10}}
\put(65,35){\vector(0,-1){0.12}} \linethickness{0.3mm}
\multiput(15,45)(0.12,-0.24){42}{\line(0,-1){0.24}}
\put(20,35){\vector(1,-2){0.12}}
\end{picture}

     \caption{Schematic plot of the mean field transition
     temperature $T_c^{MF}$ and the Kosterlitz-Thouless temperature
     $T_{KT}$ vs. $Q\xi$, where
     $Q$ is the modulating wavevector of the inhomogeneous coupling constant;
     $\xi$ is the superconducting coherence length;
     $T_{c,max}=\frac{2C}{\pi}\omega_De^{-1/N_F(\bar U+U_Q)}$ is the
     maximum $T_c^{MF}$; $T_{c,a}\equiv\frac{2C}{\pi}\omega_De^{-1/N_F\bar
     U}$ is the mean field $T_c$ for a uniform coupling $\bar U$.
      1,2,3, and 4 are labels of different regimes described in the text.
      The qualitative feature of these results on $T_c$ are similar to those of Ref. \onlinecite{Podolsky} on clean superconductors.}\label{Tc}
\end{figure}
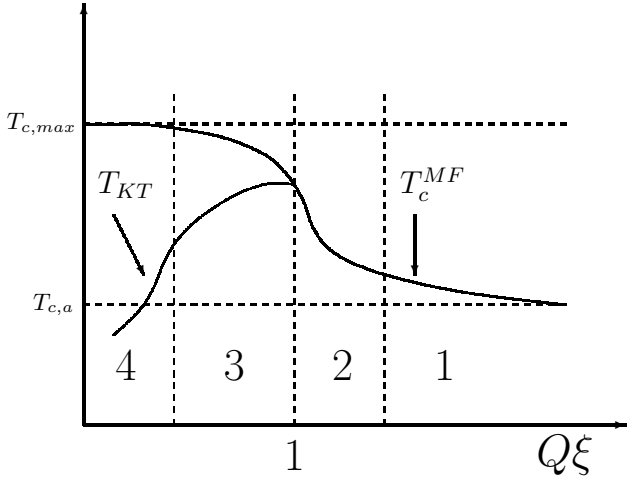

The value of $2\bar{E_g}/T_c^{MF}$ and $2\bar{E_g}/T_{KT}$ vs.
the entire range of $Q\xi$ is plotted schematically in FIG.
\ref{result}, with regimes 1-4 explicitly labeled in the graph.
Schematic results of $T_c^{MF}$ and $T_{KT}$ vs. $Q\xi$ are
summarized in FIG. \ref{Tc}.

Finally, we discuss connections with thin film experiments \cite{subgap,superweak}. A
straightforward realization of inhomogeneous coupling is in
disordered superconductor-normal-metal (SN) bilayer thin films.  In
a homogeneous bilayer SN with thickness smaller
than the coherence length $\xi$, mean field analysis yields that $T_c$ and the energy gap $E_g$ of the system are
determined by the averaged coupling constant
\cite{Cooper,deGennes,FF}
\begin{equation}\label{601}
U_{eff}=\frac{d_SN_S}{d_SN_S+d_NN_N}U,
\end{equation}
where $U$ is the pairing coupling in the superconducting layer, $d$
is the thickness, $N$ is the DOS at the Fermi energy, and the subscripts $S$
and $N$ denote the superconductor and normal metal layers
respectively. Thus the ratio $2E_{g(T=0)}/T_c$ is expected to remain
at the BCS value $2\pi/C\approx3.52$ in a homogeneous SN bilayers.
Nevertheless, from (\ref{601}) one observes that a spatially
inhomogeneous thickness $d_{S,N}(x)$ (which is also consistent with
the granular morphology of the sample\cite{morphology}) leads to a
nonuniform coupling $U(x)$ even if the original coupling $U$ is
homogeneous. Therefore thickness variation generically leads to a
superconductor with inhomogeneous pairing coupling. According to our
results, a deviation of $2E_g/T_c$ from $3.52$ is expected in such a
system.

Indeed our study was motivated by such observations. In Refs.
\onlinecite{subgap,superweak} Long et al. report measurements of
recently fabricated a series of Pb-Ag bilayer thin films, with
thickness $d_{Pb}=4$nm and $d_{Ag}$ increases from $6.7$nm to
$19.3$nm. They observed a significant reduction of
$2\bar{E_g}/T_c^{MF}$ from the expected value $\sim3.52$, where
$\bar{E_g}$ is the spatially averaged gap extracted from
tunneling measurement of the DOS, and $T_c^{MF}$ is measured as
the temperature at which $R(T)$ drops to half of its normal state
value, and the resistive transition is sharp and well-defined.
This suppression of $2\bar{E_g}/T_c^{MF}$ is more pronounced in
systems with thicker Ag thereby lower $T_c^{MF}$. In these samples
with $T_c^{MF}$ decreasing from $2.55$K to $0.72$K with increasing
$d_{Ag}$, the ratio $2\bar{E_g}/T_c^{MF}$ decreases from $\sim3.6$
to $\sim2.6$ (see FIG 3(b) of Ref. \onlinecite{superweak}).

\begin{figure}
\includegraphics[scale=0.5]{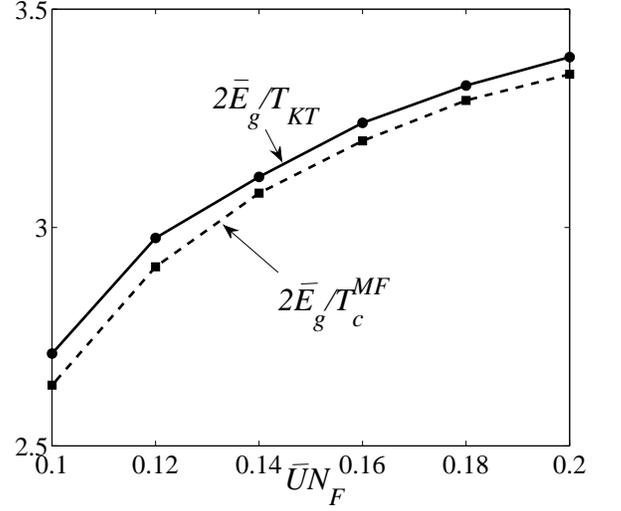}
     \caption{The ratios of the spatially averaged gap $\bar{E_g}$  to the mean field transition temperature
     $T_c^{MF}$ or the Kosterlitz-Thouless transition temperature
     $T_{KT}$ vs. $\bar{U}N_F$. $U_QN_F=0.002$, $Q\xi=0.3$.  Here,
     $\bar U$ and $U_Q$ are the uniform and oscillating components of
     the coupling constant, respectively. $N_F$ is the density of
     states of the normal state; $Q$ is the modulating wavevector of
     the inhomogeneous coupling constant; $\xi$ is the superconducting
     coherence length.  Since $T_c^{MF}$ monotonically increases with $\bar U$,
     this result resembles the experimental data of
     Ref. \onlinecite{superweak} (see FIG. 3(b) {\it ibid.}), which shows that the lower the measured
     $T_c$ of a thin-film bilayer is, the smaller the ratio
     $2E_g/T_c$.The estimated numerical error is about
     0.02. }\label{ubar}
\end{figure}

These results can be qualitatively well understood by our study.
The reduction of $2E_g/T_c^{MF}$ from $3.52$, together with the
observed fact that the resistive transition is sharp and
well-defined\cite{subgap}, implies that the experimental systems
are in the regime (2) or (3) of our theoretical results summarized
above (see FIG. \ref{result}). In these regimes both
$2\bar{E_g}/T_c^{MF}$ and $2\bar{E_g}/T_{KT}$ are lower than
$3.52$, and the phase fluctuation is either absent or small enough
to keep $T_{KT}$ close to $T_c^{MF}$, explaining the sharp
resistive transition. For samples with lower $T_c$, $\bar{U}$ is
smaller. Therefore, if we assume roughly the same amount of $U_Q$
for all samples, the effect of inhomogeneity will be stronger for
samples with lower $T_c$ samples, and, consequently, the
gap-to-$T_c$ ratio is even smaller for them. To make a rough
comparison, we have calculated the gap-$T_c$ ratio vs. $\bar U$
for fixed $U_Q$ and plotted the results in FIG. \ref{ubar}.
Although not claiming more than a qualitative explanation of the
bilayer measurements, we note that our FIG. \ref{ubar} resembles
FIG. 3(b) of Ref. \onlinecite{superweak}.

An interesting venue for future research, which may extend to more 2d
superconducting systems, is to consider a general fluctuation of the
pairing interaction, not restricted to a particular wave number, but
rather having a particular correlation length. In addition, aside from the low gap-$T_c$ ratio, Ref. \onlinecite{subgap} has also reported an unexpected subgap density of states of quasiparticles in the same bilayer materials. Although our current model does not produce this behavior, one expects that it could be explained by including large spatial fluctuations of the pairing interaction (e.g. $\frac{U_Q}{\bar{U}}\sim1$), which strongly suppress the gap, and the effect of mesoscopic fluctuations which tend to produce subgap states\cite{Simons2001}.

\acknowledgments We would like to thank D. Podolsky for several
enlightening discussions. The work of IK was supported in part by the National Science Foundation
under Grant No. PHY05-51164.

\appendix

\section{Calculation of $\Delta_{(T=0)}$ in the limit $Q\xi\gg1$ \label{appA}}
Here we show some calculation details in deriving equation
(\ref{31}). At $T=0$ the self-consistency equations are
\begin{equation}\label{25}
\Delta_0=N_F\bar
U\left(\int_0^{\omega_D}d\omega\sin\theta_0\right)+\frac{N_FU_Q}2\left(\int_0^{\omega_D}d\omega\theta_1\cos\theta_0\right);\nonumber
\end{equation}
\begin{equation}\label{26}
\frac{\Delta_1}2= \frac{N_F\bar
U}2\left(\int_0^{\omega_D}d\omega\theta_1\cos\theta_0\right)
+\frac{N_FU_Q}2\left(\int_0^{\omega_D}d\omega\sin\theta_0\right).\nonumber
\end{equation}
The evaluation of the integrals gives (define
$a=\frac{DQ^2/2}{\Delta_0}$ and $x_0=\omega_D/\Delta_0$):
\begin{equation}\label{27}
\int_0^{\omega_D}d\omega\sin\theta_0=\Delta_0\arcsinh\left(\frac{\omega_D}{\Delta_0}\right)\approx\Delta_0\ln\left(\frac{2\omega_D}{\Delta_0}\right);\nonumber
\end{equation}

\begin{eqnarray}\label{28}
\int_0^{\omega_D}d\omega\theta_1\cos\theta_0
&=&\frac{\Delta_1}{2a}\left\{-2\arctan(x_0)+2a\arcsinh(x_0)\right.\nonumber\\
&-&\sqrt{a^2-1}\left[\arctanh\left(\frac{x_0\sqrt{a^2-1}+1}{a\sqrt{x_0^2+1}}\right)\right.\nonumber\\&+&\arctanh\left(\frac{x_0\sqrt{a^2-1}-1}{a\sqrt{x_0^2+1}}\right)
\nonumber\\&-&\left.\left.2\arctanh\left(\frac{x_0}{\sqrt{a^2-1}}\right)\right]\right\}.
\end{eqnarray}
We take the limit $x_0=\frac{\omega_D}{\Delta_0}\gg1$ and
$a=(Q\xi)^2\gg1$ simultaneously, but their relative ratio might be
either large or small. Also using $\arctanh(z)=1/2\ln(|1+z|/|1-z|)$,
one can show that in this limit the above integral equals
\begin{eqnarray}\label{35}
&=&\frac{\Delta_1}{2a}\left\{2a\ln(2x_0)-a\left[\frac12\ln\left(\frac{2x_0a}{\frac{a}{2x_0}+\frac{x_0}{2a}-1}\right)\right.\right.\nonumber\\&+&\left.\left.\frac12\ln\left(\frac{2x_0a}{\frac{a}{2x_0}+\frac{x_0}{2a}+1}\right)
+\ln\left(\frac{|x_0-a|}{x_0+a}\right)\right]\right\}\nonumber\\
&=&\frac{\Delta_1}{2}\left\{2\ln(2x_0)-\left[\ln\left(\frac{2x_0a}{|\frac{x_0}{2a}-\frac{a}{2x_0}|}\right)
+\ln\left(\frac{|x_0-a|}{x_0+a}\right)\right]\right\}\nonumber\\
&=&\Delta_1\ln\left(1+\frac{x_0}a\right)=\Delta_1\ln\left(1+\frac{2\omega_D}{DQ^2}\right)=\frac{\Delta_1}{\bar
UN_F}K_1,\nonumber
\end{eqnarray}
where $K_1$ has exactly the same form as defined in (\ref{20}).

\bibliography{reference}

\end{document}